\documentclass{PoS}
\usepackage{graphicx}
\usepackage{lineno}
\usepackage{subfigure}
\usepackage{here}

\title{Observing ultra-high energy cosmic rays with prototypes of the Fluorescence detector Array of Single-pixel Telescopes (FAST) in both hemispheres}

\ShortTitle{Observing UHECRs with prototypes of FAST in both hemispheres}

\author{ \speaker{Toshihiro Fujii}$^{,a}$, Max Malacari$^{b}$, John Farmer$^{b}$, Justin Albury$^{c}$, Jose A. Bellido$^{c}$, Ladislav Chytka$^{d}$, Petr Hamal$^{d}$, Pavel Horvath$^{e}$, Miroslav Hrabovsky$^{e}$, Jiri Kvita$^{e}$, Dusan Mandat$^{d}$, Massimo Mastrodicasa$^{f}$, John N. Matthews$^{g}$, Stanislav Michal$^{d}$, Xiaochen Ni$^{b}$, Libor Nozka$^{d}$, Miroslav Palatka$^{d}$, Miroslav Pech$^{d}$, Paolo Privitera$^{b}$, Petr Schovanek$^{d}$, Francesco Salamida$^{f}$, Radomir Smida$^{b}$, Stan B. Thomas$^{g}$, Petr Travnicek$^{d}$, Martin Vacula$^{e}$  (FAST Collaboration)\footnote{https://www.fast-project.org}\\
$^{a}$Hakubi Center for Advanced Research and Graduate School of Science, Kyoto University, Sakyo, Kyoto, Japan\\
$^{b}$Kavli Institute for Cosmological Physics, University of Chicago, Chicago, IL, USA\\
$^{c}$Department of Physics, University of Adelaide, Adelaide, S.A., Australia\\
$^{d}$Institute of Physics of the Academy of Sciences of the Czech Republic, Prague, Czech Republic\\
$^{e}$Palacky University, Joint Laboratory of Optics, Olomouc, Czech Republic\\
$^{f}$Department of Physical and Chemical Sciences, University of L'Aquila  and INFN LNGS\\
$^{g}$High Energy Astrophysics Institute and Department of Physics and Astronomy, University of Utah, Salt Lake City, UT, USA\\
E-mail: \href{mailto:fujii@cr.scphys.kyoto-u.ac.jp}{\rm fujii@cr.scphys.kyoto-u.ac.jp}\\
}

\abstract{The origin and nature of ultra-high energy cosmic rays (UHECRs) are hot topics in the astroparticle physics community. 
The Fluorescence detector Array of Single-pixel Telescopes (FAST) is a design for a next-generation ground-based UHECR observatory, addressing the requirements for a large-area, low-cost detector suitable for measuring the properties of the highest energy cosmic rays with an unprecedented aperture. 
We have developed a full-scale prototype consisting of four 200 mm photomultiplier tubes at the focus of a segmented mirror of 1.6 m in diameter. 
Over the last three years, we have installed three prototypes at the Telescope Array Experiment in Utah, USA. 
These telescopes have been steadily taking data since installation. 
We report on preliminary results of the full-scale FAST prototypes, including measurements of UHECRs, and distant ultra-violet lasers used to study the atmospheric transparency. 
Furthermore, we discuss the installation of an additional identical FAST prototype at the Pierre Auger Observatory in Argentina. 
Possible benefits to the Telescope Array Experiment and the Pierre Auger Observatory include a comparison of the transparency of the atmosphere above both experiments, a study of the systematic uncertainty associated with their existing fluorescence detectors, and a cross-calibration of their energy and $X_{\max}$ scales.}

\FullConference{36th International Cosmic Ray Conference -ICRC2019-\\
		July 24th - August 1st, 2019\\
		Madison, WI, U.S.A.}

\begin{document}

\section{Introduction of ultra-high energy cosmic rays}
\label{intro}
Cosmic rays with energies above 10$^{19}$\,eV, so-called ultra-high energy cosmic rays (UHECRs), are the highest energy particles in the universe. Their origins are ostensibly related to extremely energetic astrophysical phenomena, such as gamma-ray bursts, active galactic nuclei, or other exotic processes such as the decay or annihilation of super-heavy relic particles created in an early phase of the development of the universe~\cite{Hillas:1985is}.
However, their origin and acceleration mechanism at the highest energies are still largely unknown, making them one of the most intriguing and important mysteries in particle astrophysics and astronomy. 

Following the detection of a cosmic ray event with an energy of 10$^{20}$\,eV by J. Linsley~\cite{Linsley:1963km} in 1963, K. Greisen, G.\,T. Zatsepin and V.\,A. Kuzmin predicted the UHECR energy spectrum to be suppressed above 10$^{19.7}$\,eV due to the interaction of high-energy particles with the 2.7\,K cosmic microwave background radiation via pion production, the so-called GZK cutoff~\cite{bib:gzk1, bib:gzk2}. 
If the GZK cutoff exists, the origin of UHECRs is significantly restricted to nearby sources distributed non-uniformly within 50-100\,Mpc. 
Additionally, UHECRs propagate with less deflection by magnetic fields due to their enormous kinetic energies. 
As a result, the arrival directions of UHECRs should be correlated with the directions of extremely energetic sources or objects, leading to the possibility of next-generation particle astronomy with UHECRs, illuminating extremely energetic phenomena in the nearby universe.

\section{Detection methods and recent results from current observatories}
\label{results}

Given the minute flux of UHECRs, less than one particle per century per square kilometer at the highest energies, a very large area must be instrumented in order to collect significant statistics.
The energy, arrival direction, and mass composition of UHECRs can be inferred from studies of the cascades of secondary particles (Extensive Air Shower, EAS) produced by their interaction with the Earth's atmosphere. 

Two well-established techniques are used for UHECR detection: arrays of detectors (e.g. plastic scintillators or water-Cherenkov stations) sample EAS particles reaching the ground, and large-field-of-view telescopes allow for reconstruction of the shower development in the atmosphere by imaging ultra-violet fluorescence light from atmospheric nitrogen excited by EAS particles.
Two giant observatories, one in each hemisphere, the Pierre Auger Observatory (Auger) in Mendoza, Argentina~\cite{bib:auger} and the Telescope Array Experiment (TA) in Utah, USA~\cite{bib:tafd, bib:tasd}, combine the two techniques with arrays of particle detectors overlooked by fluorescence detectors (FDs).

Both the Auger and TA observatories have measured a dip in the energy spectrum at around 10$^{18.7}$\,eV, and a suppression above 10$^{19.7}$\,eV~\cite{Verzi:2017hro}. 
The energy of the suppression is consistent with expectation from the GZK cutoff, with the shape of the suppression being discrepant between the two measurements with the Auger spectrum softer than the TA measurement. 
The mass composition reported by Auger through $X_{\max}$ (the depth in the atmosphere at which the EAS reaches its maximum energy deposit) suggests a transition from light nuclei at around 10$^{18.3}$\,eV to heavier nuclei up to energies of 10$^{19.6}$\,eV~\cite{bib:mass_auger, bib:mass_implication_auger}. The mass composition reported by TA is consistent with a light composition above 10$^{18.2}$ eV~\cite{bib:ta_composition}. The $X_{\max}$ distributions measured by Auger and TA below 10$^{19}$\,eV are consistent within systematic uncertainties~\cite{deSouza:2017wgx}.

In 2017, Auger reported on a measured anisotropy above 8\,EeV\footnote{1\,EeV = 10$^{18}$\,eV}, indicating an obvious dipole structure of 6.5\% amplitude with a 5.2$\sigma$ significance~\cite{bib:dipole_auger}, which supports an extragalactic origin for these ultra-high energy particles.
An enhancement of the dipole amplitude above 4\,EeV is also reported~\cite{Aab:2018mmi}.
TA has measured an excess of cosmic rays above 57\,EeV as a ``hotspot'' centered at a right ascension of 147$^{\circ}$ and a declination of 43$^{\circ}$ with a 3.4$\sigma$ significance~\cite{bib:hotspot_ta}. 
Auger has reported a 4.0$\sigma$ correlation between the positions of nearby starburst galaxies and the arrival directions of 9.7\% of their measured UHECR events above 39\,EeV~\cite{bib:sbg_auger}.

Except for the Auger dipole result above 8\,EeV, these inconclusive results at the highest energies are limited by statistics at the highest energies due to the flux suppression. 
To further advance and establish the field of charged particle astronomy, a future ground array will require an unprecedented aperture, which is larger by an order of magnitude, and mass composition sensitivity above 100\,EeV.
In order to cover this enormous target volume, future detectors should be low-cost and easily-deployable. 
A worldwide collaboration is essential in successfully building such a giant ground array.

\begin{figure}
\centering
\includegraphics[width=1.0\linewidth,clip]{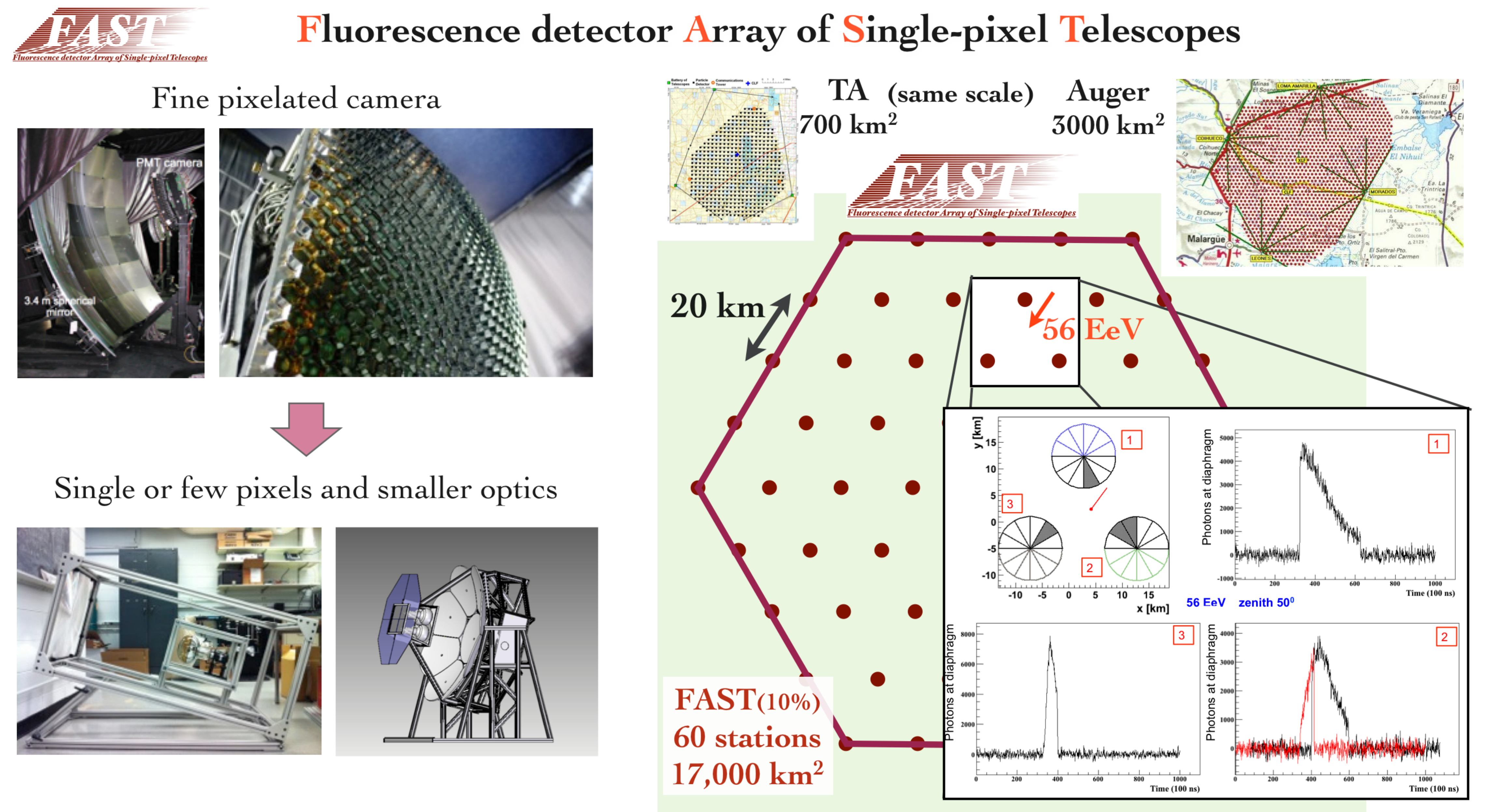}
\caption{A possible schematic view of the Fluorescence detector Array of Single-pixel Telescopes: one of the possible solutions for a future giant ground array~\cite{bib:fast}. As a reference, the ground coverage of TA and Auger make up only 10\% of the ground coverage of a proposed full-scale array of FAST detectors. The expected signal emitted from a UHECR with an energy of 56\,EeV is shown in a simulation. }
\label{bib:fast}
\end{figure}

\section{Fluorescence detector Array of Single-pixel Telescopes (FAST)} 
One of the possible solutions capable of fulfilling these requirements is a ground-based fluorescence detector array that is low-cost, easily-deployable, and has an unprecedented aperture which is an order of magnitude larger than that of current UHECR observatories.

The Fluorescence detector Array of Single-pixel Telescopes (FAST)\footnote{https://www.fast-project.org} consists of compact FD telescopes featuring a smaller light collecting area and far fewer pixels than current generation FD designs, leading to a significant reduction in cost.
In the FAST design, a 30$^{\circ}$ $\times$ 30$^{\circ}$ field-of-view is covered by four 200\,mm photomultiplier-tubes (PMTs) at the focal plane of a compact segmented mirror of 1.6\,m in diameter~\cite{bib:fast_optics}. 
A significant cost reduction is expected due to the compact design of FAST, with smaller light collecting optics, a smaller telescope housing, and a smaller number of PMTs and associated electronics. 
Each FAST station would consist of 12 telescopes, covering 360$^{\circ}$ in azimuth and 30$^{\circ}$ in elevation. 
Powered by solar panels and with wireless connection, they would be deployed in a triangular arrangement with a 20\,km spacing. 

Figure~\ref{bib:fast} shows a possible schematic view of FAST, and the expected signal from a UHECR shower with coincidence detections at three adjacent stations.
An example of the ground coverage using 60 stations is also indicated with a comparison to Auger and TA coverages. 
To achieve an order of magnitude larger effective aperture than current observatories, 500 stations are required after accounting for the standard FD duty-cycle.
The operation of a full-size FAST array will provide conclusive results on the origin and acceleration mechanism towards 100\,EeV, and open the window to charged-particle astronomy using UHECRs.

\begin{figure}[tb]
  \centering
  \subfigure{\includegraphics[width=0.9\linewidth]{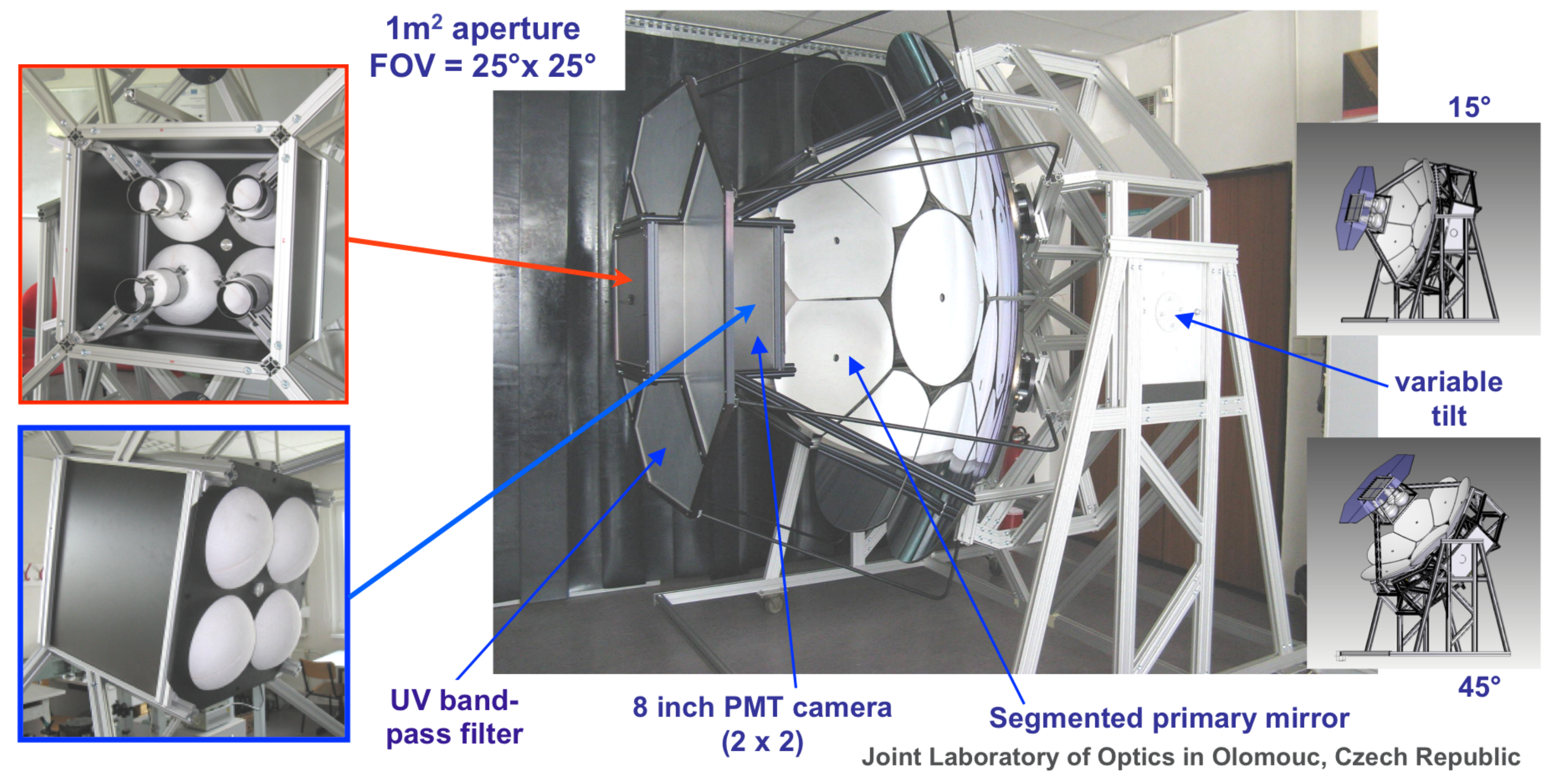}}
  \subfigure{\includegraphics[width=0.9\linewidth]{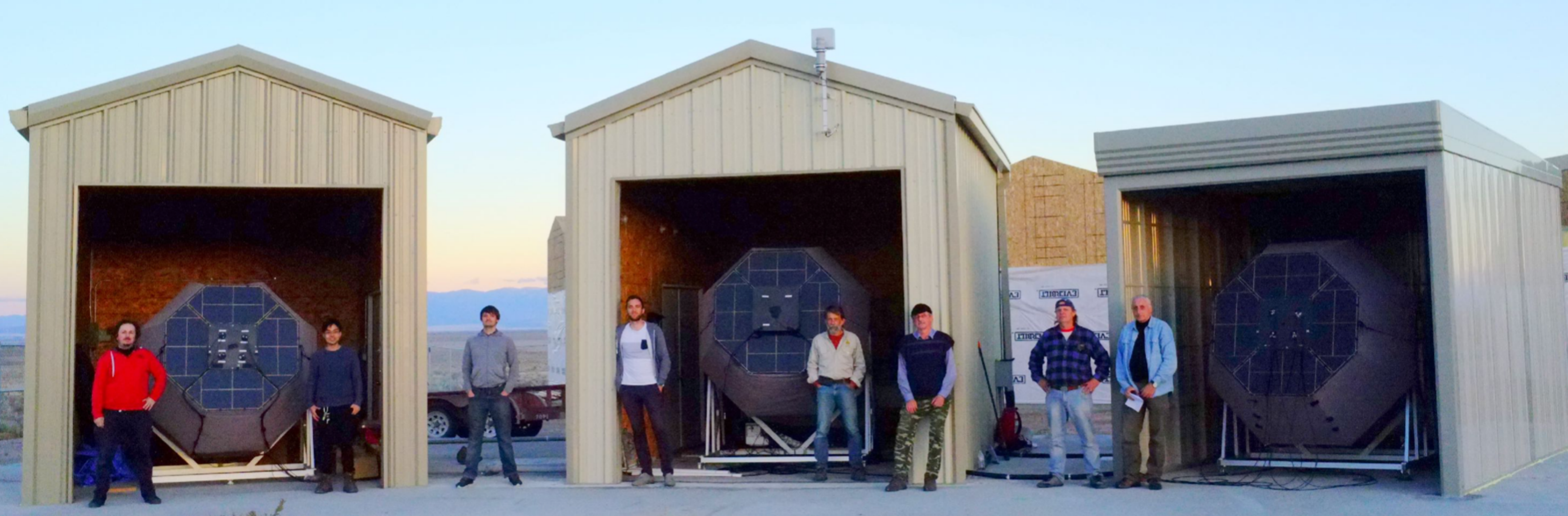}}
  \caption{The design of the full-scale FAST prototype and the three FAST prototypes installed at the Black Rock Mesa site of the Telescope Array Experiment.}
  \label{fig:fast_tel}  
\end{figure}
Motivated by detections of UHECRs using a single 200\,mm PMT at the focus of a 1\,m$^2$ Fresnel lens system~\cite{bib:fast}, three full-scale FAST prototypes have been installed at TA site as shown in Figure~\ref{fig:fast_tel}. 
The combined field of view of the prototypes covers 30$^{\circ}$ in elevation and 90$^{\circ}$ in azimuth.
The telescope frames were assembled on site, before the PMTs were mounted in the camera box and the ultra-violet band-pass filter was installed at the telescope aperture.
The telescopes were aligned astrometrically using a camera mounted to the exterior of the frame~\cite{bib:fast_optics}. 
Following their installation, stable operation of the prototypes began via a remote connection, utilizing external air shower triggers from the adjacent TA fluorescence detector.

One of the three FAST prototypes observes a half-hourly vertical laser signal at a distance of 21\,km. 
The laser signal is useful for monitoring the atmospheric transparency above the detector, which is one of the largest systematic uncertainties in the analysis of data collected using the fluorescence technique. 
An optical camera with a fish-eye lens and photo-sensor have also been installed for monitoring of the night-sky background and cloud coverage~\cite{Mandat:2019qud}.


\begin{figure}
  \centering
  \subfigure{\includegraphics[width=0.32\linewidth]{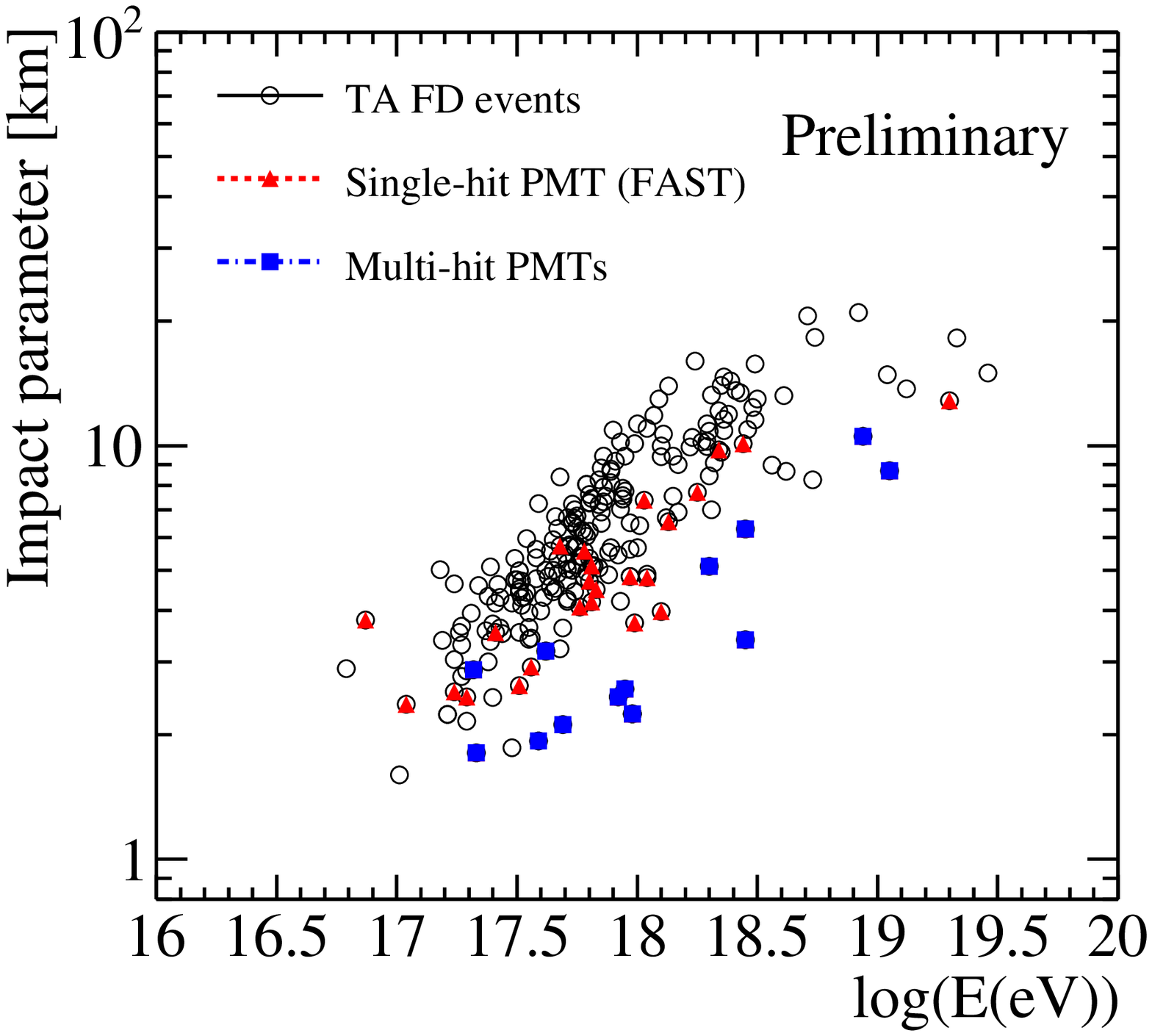}}
  \subfigure{\includegraphics[width=0.32\linewidth]{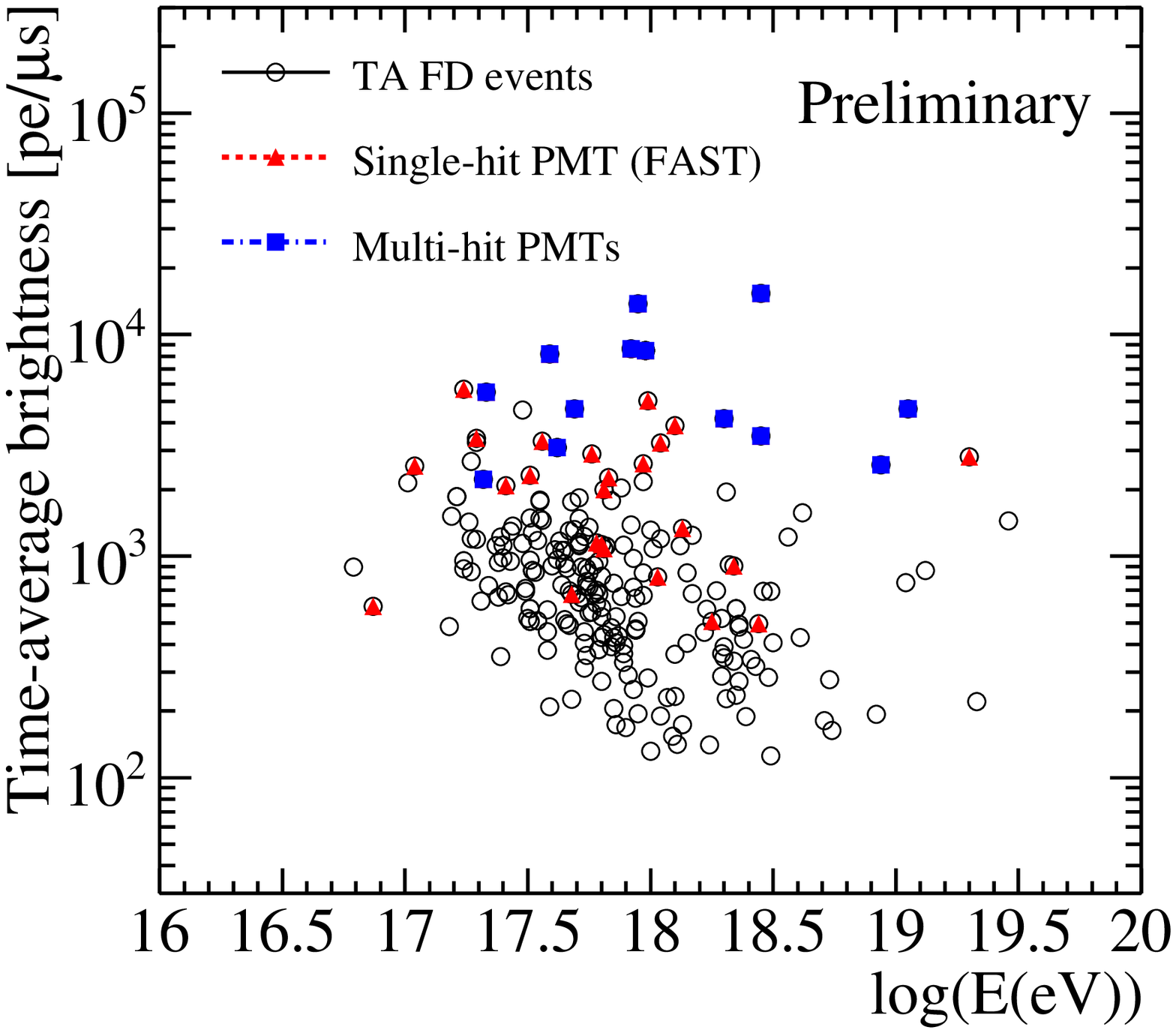}}
  \subfigure{\includegraphics[width=0.32\linewidth]{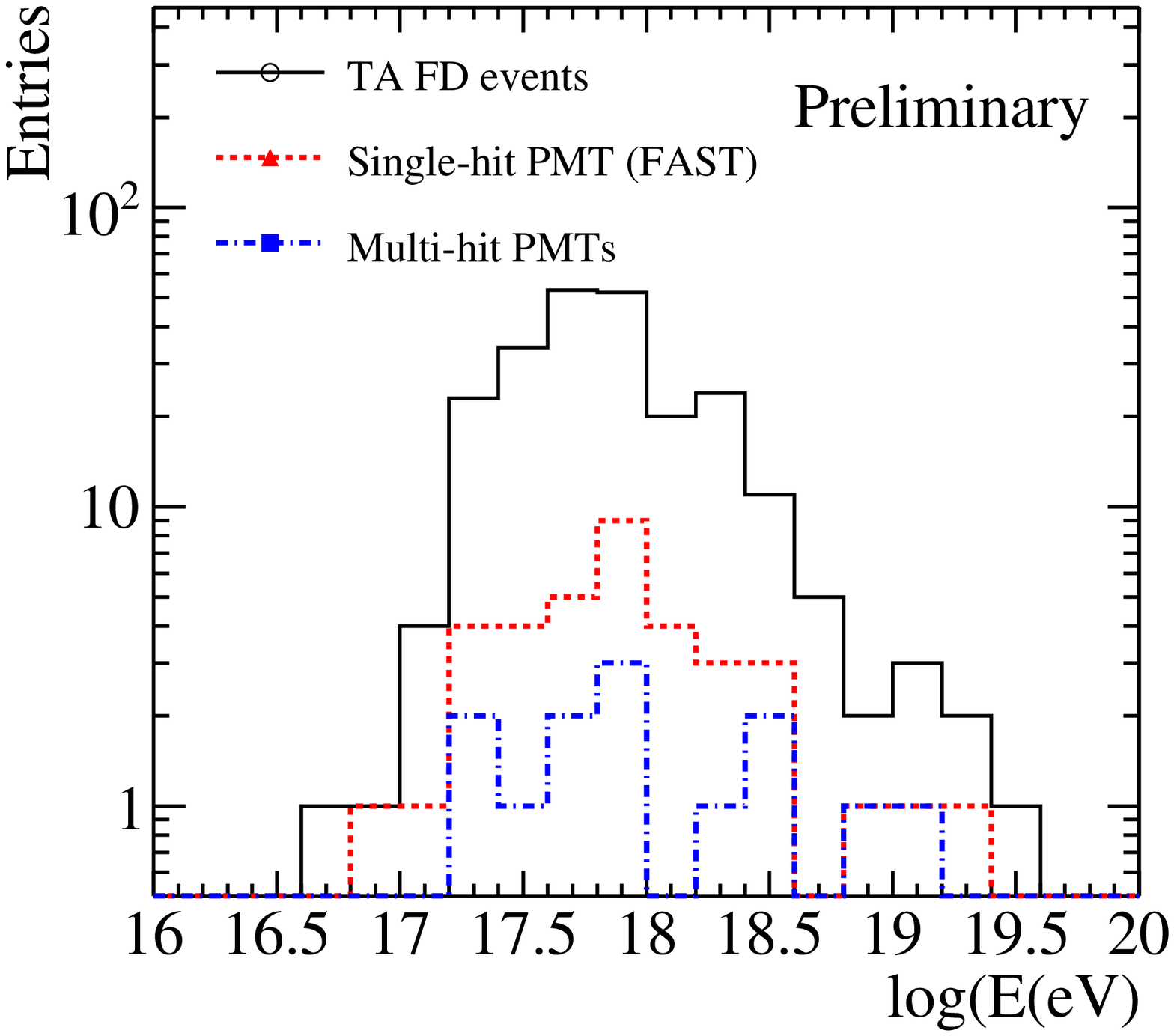}}
  \caption{The impact parameter and time-average brightness of detected the EAS as a function of the reconstructed energy, along with an energy histogram. These parameters are reconstructed by the TA FD. Open-circles indicate all TA FD reconstructed events, red triangles and blue squares show single-hit PMT and multi-hit PMT events by the FAST prototypes, respectively.}
  \label{fig:distribution}  
\end{figure}

\section{Preliminary results using three FAST prototypes at Telescope Array Experiment}
As of June 2019, the total observation time of the FAST prototypes is 545 hours. 
A period from the 6$^{\textrm{th}}$ of October 2018 to the 14$^{\textrm{th}}$ of January 2019, corresponding to 52 hours during which all three prototypes were operational was used to search for coincident EAS detections between the TA FD and the FAST prototypes.
236 showers were found in this period with corresponding monocular reconstructions with the TA FD~\cite{bib:tafd_spectrum2016}. 
We searched for significant signals, defined as a $\ge 6\sigma$ signal-to-noise ratio over $\ge500$ nanoseconds signal duration, in time coincidence with detections by the TA FD.
We found 37 out of the 236 EAS to have significant signals as measured by the FAST prototypes, with 13 EAS having a significant signal in more than one PMT.

\begin{figure}
  \centering
  \subfigure{\includegraphics[width=0.85\linewidth]{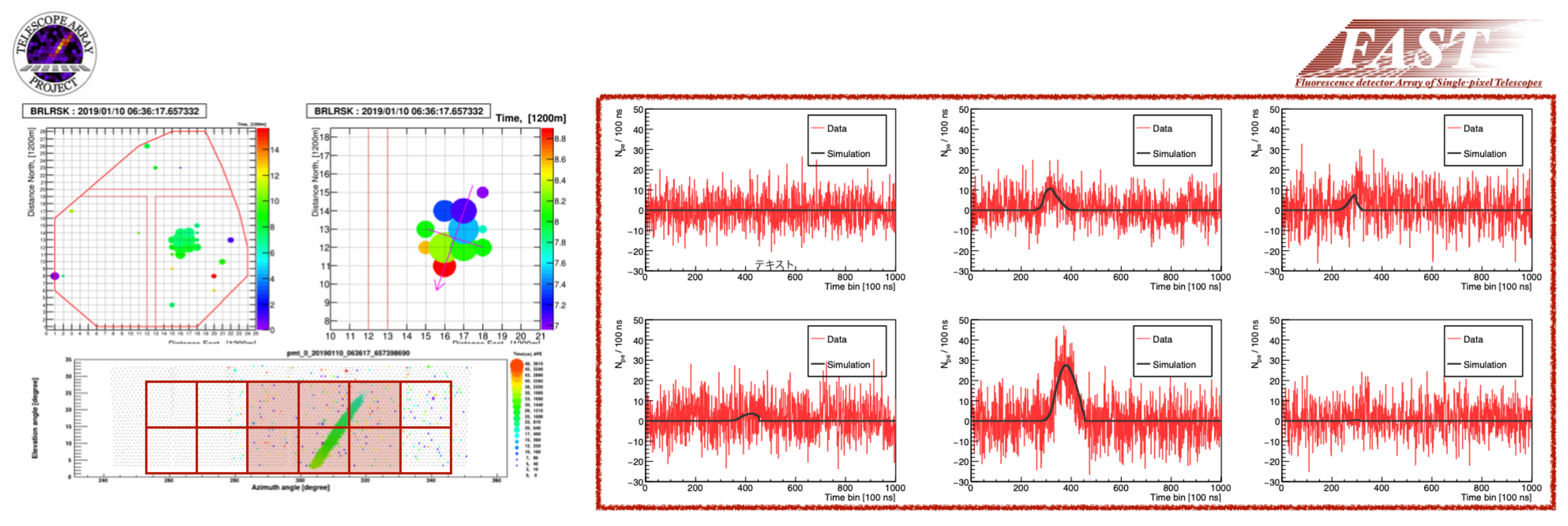}}
  \subfigure{\includegraphics[width=0.65\linewidth]{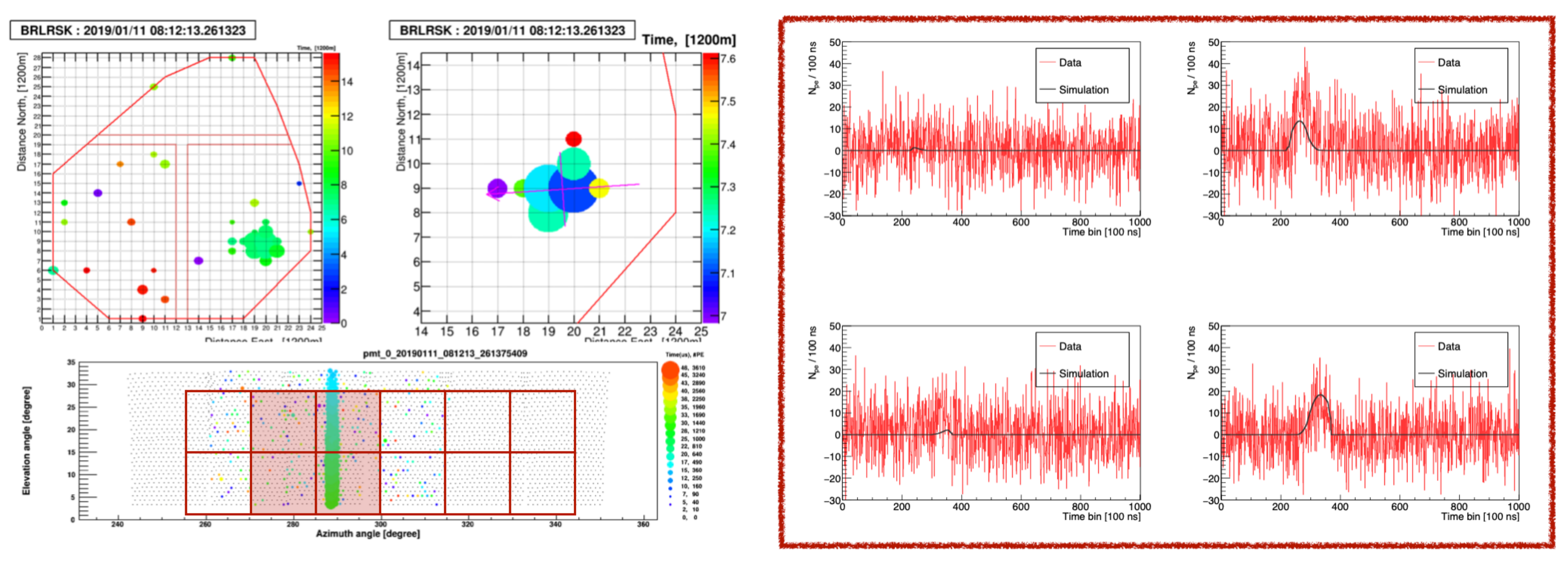}}
  \caption{The two highest energy cosmic rays detected with TA detectors and FAST prototypes. TA displays (left) show event data measured by the surface detector array, and the PMT signals measured by the TA fluorescence telescopes. The size of the circles indicate the signal amplitude and the color represents the signal timing. The red square grid corresponds to individual PMT field-of-views of FAST prototypes. The waveforms (right) observed with the FAST prototypes. The red histograms show the recorded data and the black curves indicate the best-fit signal from a top-down reconstruction.}
  \label{fig:event}
\end{figure}
Figure~\ref{fig:distribution} shows the impact parameter and time-average brightness of the detected EAS as a function of energy, along with an energy histogram. These parameters are reconstructed by the TA FD.
The maximum detectable impact parameter is approximately 20\,km at 10$^{19.5}$\,eV, with brighter signal showers being detected by more than one PMT on average. 
Two cosmic rays with energies above 10\,EeV were detected with the three prototypes during the 52\,hours observation time, suggesting an expected event rate of $\sim$25\,events per year if a 15\% duty cycle is assumed.

A ``top-down'' reconstruction algorithm for data collected by the FAST prototypes has been implemented to determine the best-fit shower parameters by comparing the measured signal trace to a library of simulated templates.
Figure~\ref{fig:event} shows event displays of the two highest energy cosmic rays detected by the TA FD and FAST prototypes, along with a comparison between the preliminary waveforms of the FAST prototypes with simulated signals from the result of the top-down reconstruction.
The preliminary energy and $X_{\max}$ values reconstructed by the top-down method using FAST data are 19\,EeV and 808\,g/cm$^2$, and 10\,EeV and 830\,g/cm$^2$, respectively.
The simulated waveforms corresponding to these parameters show reasonable agreement with the data, although further understanding of the telescope calibration factors are required to reduce the discrepancy.

\section{Installation of a FAST prototype at the Pierre Auger Observatory} 
\begin{figure}
  \centering
  \includegraphics[width=1.0\linewidth]{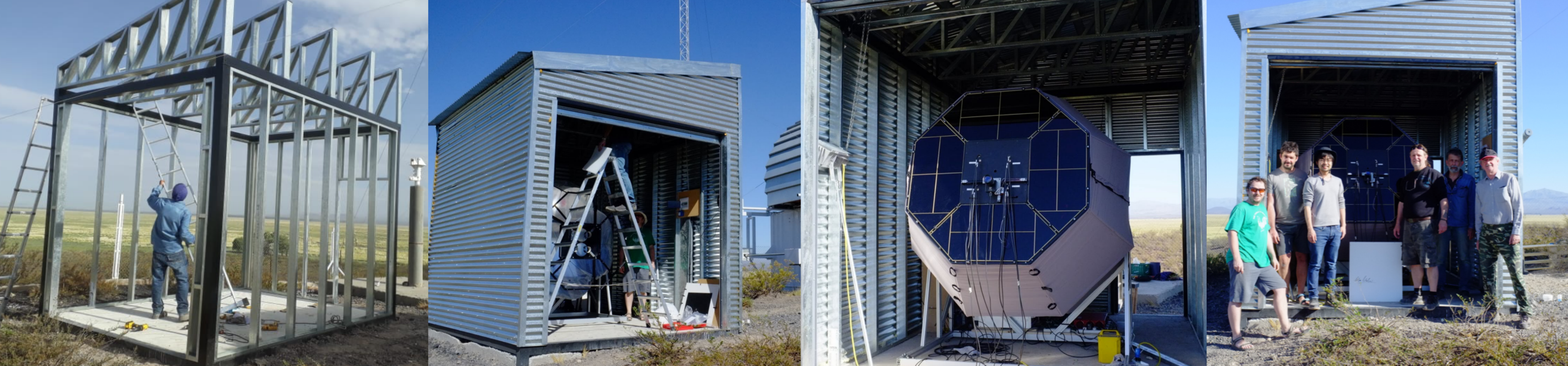}
  \caption{Photographs of the FAST prototype installed at the Pierre Auger Observatory.}
  \label{fig:auger}
\end{figure}
We installed an additional identical FAST prototype at the Auger site in April 2019 and began a remote operation. Figure~\ref{fig:auger} shows photographs taken during the installation. 
Signals from a distant laser, along with Cherenkov-dominated signals from close-by UHECR showers have already been observed with this new prototype. The total observation time is 85\,hours as of June 2019. 
This identical FAST prototype will allow for a cross-calibration of the energy and $X_{\max}$ scales of Auger and TA, as well as a comparison between the atmospheric transparency at both sites, an important source of systematic uncertainty in the fluorescence technique.

\section{Summary}
To achieve a future 30,000\,km$^{2}$ effective ground coverage, we have developed a low-cost, easily-deployable fluorescence detector optimized for the detections of the highest energy cosmic rays. 
Three FAST prototypes have been installed at the Telescope Array Experiment, and one prototype installed at the Pierre Auger Observatory. 
We have achieved stable observation of UHECRs in both hemispheres, with a future goal to perform a cross-calibration of the TA and Auger energy and $X_{\max}$ scales.
We will continue the steady remote operation of all four FAST prototypes in order to further increase the UHECR statistics in both hemispheres. 

\section*{Acknowledgements}
This work was supported by JSPS KAKENHI Grant Number 18KK0381, 18H01225, 15H05443,
and Grant-in-Aid for JSPS Research Fellow 16J04564 and JSPS Fellowships H25-339, H28-4564.
This work was supported by a research fund from the Hakubi Center for Advanced Research, Kyoto University.
This work was partially carried out by the joint research program of
the Institute for Cosmic Ray Research (ICRR) at the University of Tokyo.
This work was supported in part by NSF grant PHY-1713764, PHY-1412261 and by the Kavli
Institute for Cosmological Physics at the University of Chicago through
grant NSF PHY-1125897 and an endowment from the Kavli Foundation and its founder Fred Kavli.
The Czech authors gratefully acknowledge the support of the Ministry of Education,
Youth and Sports of the Czech Republic project No. LTAUSA17078, CZ.02.1.01/0.0/17\_049/0008422, LTT 18004.
The Australian authors acknowledge the support of the Australian Research Council, through Discovery Project DP150101622.
The authors thank the Pierre Auger and Telescope Array Collaborations for providing logistic support 
and part of the instrumentation to perform the FAST prototype measurement and for productive discussions.
\vspace{0.1cm}

\bibliography{fast_icrc2019}
\bibliographystyle{JHEP}

\end{document}